\documentclass[12pt]{article}
\usepackage{amsfonts,amssymb}
\usepackage{color}
\usepackage{epic}
\usepackage{graphics}

\def\lddots{\mathinner{\mkern1mu\raise1pt\hbox{.}\mkern2mu
\raise4pt\hbox{.}\mkern2mu\raise7pt\vbox{\kern7pt\hbox{.}}\mkern1mu}}
\makeatletter
\def\numberbysection{\@addtoreset{equation}{section}
\def\theequation{\thesection.\arabic{equation}}}
\makeatother

\numberbysection

\newcommand{\be}{\begin{eqnarray}}
\newcommand{\ee}{\end{eqnarray}}
\newcommand{\non}{\nonumber}

\begin{document}

\begin{titlepage}
\vskip 0.4cm \strut\hfill \vskip 0.8cm
\begin{center}

{\bf {\Large $A_n^{(1)}$ affine Toda field theories with
integrable boundary conditions revisited}}

\vspace{10mm}

{\large {\bf Anastasia Doikou}}\footnote{e-mail: doikou@bo.infn.it, adoikou@upatras.gr}

\vspace{10mm}

University of Bologna, Physics Department, INFN Section \\
Via Irnerio 46, Bologna 40126, Italy

\end{center}

\vfill

\begin{abstract}

Generic classically integrable boundary conditions for the
$A_{n}^{(1)}$ affine Toda field theories (ATFT) are investigated.
The present analysis rests primarily on the underlying algebra,
defined by the classical version of the reflection equation. We use
as a prototype example the first non-trivial model of the hierarchy
i.e. the $A_2^{(1)}$ ATFT, however our results may be generalized
for any $A_{n}^{(1)}$ ($n>1$). We assume here two distinct types of
boundary conditions called some times soliton preserving (SP), and
soliton non-preserving (SNP) associated to two distinct algebras,
i.e. the reflection algebra and the ($q$) twisted Yangian
respectively. The boundary local integrals of motion are then
systematically extracted from the asymptotic expansion of the
associated transfer matrix. In the case of SNP boundary conditions
we recover previously known results. The other type of boundary
conditions (SP), associated to the reflection algebra, are novel in
this context and lead to a different set of conserved quantities
that depend on free boundary parameters. It also turns out that the
number of local integrals of motions for SP boundary conditions is
`double' compared to those of the SNP case.

\end{abstract}

\vfill \baselineskip=16pt

\end{titlepage}

\section{Introduction}

Integrability in the bulk has admittedly attracted a great deal of
research interest in recent years, however after the seminal works
of \cite{cherednik, sklyanin, cardy} particular emphasis has been
given on the issue of incorporating consistent boundary conditions
in integrable models. This shed new light into the bulk theories
themselves, and also opened the path to new mathematical concepts
and physical applications. In a more general setting the
investigation of both classical and quantum integrable systems,
particularly those with non-trivial boundary conditions, turns out
to be quite significant especially after the recent advances
within the AdS/CFT correspondence \cite{malda} uncovering the
important role of integrability \cite{miza}. A crucial question
within this frame is what would the physical implications be in
both gauge and string theories once non-trivial consistent
boundary conditions, especially the ones that may modify the bulk
behavior, are imposed to the associated lattice and continuum
integrable models (for some recent results see \cite{maldah} and
references therein). Therefore studies concerning the existence of
consistent boundary conditions that preserve integrability are of
particular significance and timeliness not only for the integrable
systems themselves, but for other active research fields.

The central purpose of the present article is the investigation of
classical integrable models when general boundaries that preserve
integrability are implemented. Among the various classes of
integrable models we choose to consider here a particular class that
is the affine Toda field theories (ATFT) \cite{mikh, olive}. The
prototype model of this class is the sine-Gordon model, which has
been extensively studied both in the bulk \cite{ZZ} as well as in
the presence of non-trivial integrable boundary conditions
\cite{GZ}. Generic affine Toda field theories with classical
integrable boundary conditions were first analyzed more than a
decade ago in \cite{durham}. A different point of view, although
regarding the same class of boundary conditions\footnote{by `same
class of boundary conditions' we mean that in both studies
\cite{durham, baz} a common underlying algebra --(classical)
$q$-twisted Yangian-- is implicitly assumed. Note however that the
analysis in \cite{durham} is classical while in \cite{baz} is
quantum.} analyzed in \cite{durham}, is presented in \cite{baz}.
Specifically, in \cite{baz} the $A_2^{(1)}$ ATFT with `dynamical'
boundary conditions --that is a quantum mechanical system is
attached at the boundary--- is investigated. Further studies
regarding the boundary ATFT at both classical and quantum level may
be also found in various articles (see e.g.
\cite{friko}--\cite{dema}).

Although the analysis in \cite{durham} seems quite exhaustive it
turns out that in simply-laced ATFT a whole class of consistent
boundary conditions is absent. Our main objective here is to
systematically search for all possible boundary conditions in
$A_n^{(1)}$ ATFT and eventually implement the missing ones. More
precisely, we assume two distinct types of boundary conditions
called soliton preserving (SP), and soliton non-preserving (SNP)
associated to two distinct algebras, i.e. the reflection algebra
\cite{sklyanin} and the twisted Yangian \cite{molev, moras}
respectively (see also relevant studies in \cite{dema, ann1, ann2,
doikoun, doikouy, crdo}).

Depending on the choice of boundary conditions certain physical
behavior is entailed. Specifically, in the context of imaginary
$A_{n}^{(1)}$ ATFT the boundary conditions introduced in
\cite{durham}, known as SNP, oblige a soliton to reflect to an
anti-soliton. In real $A_{n}^{(1)}$ ATFT on the other hand such
boundary conditions lead to the reflection of a fundamental
particle to itself. Recall that fundamental particles in real ATFT
are equivalent to the lightest bound states (breathers) of the
imaginary theory provided that $\beta \to i \beta$ ($\beta$ is the
coupling constant of the theory). It is however clear that another
possibility arises, that is the implementation of certain boundary
conditions that lead to the reflection of a soliton to itself in
imaginary ATFT or to the reflection of a fundamental particle to
its conjugate in real ATFT. These boundary conditions are known as
soliton preserving and have been extensively analyzed in the frame
of integrable quantum spin chains \cite{ann1, ann2},
\cite{dvg}--\cite{masa}.

Notwithstanding SP boundary conditions are somehow the obvious ones
in the framework of integrable lattice models they have remained
elusive in the context of $A_n^{(1)}$ ATFT for quite a long time.
Note however that in quantum spin chains in addition to the well
studied SP boundaries SNP boundary conditions were first introduced
in \cite{doikousnp} and further analyzed and generalized in
\cite{ann1, ann2, doikouy, crdo}. It is thus our primary objective
here to complete the study of integrable boundary conditions in ATFT
by introducing and fully analyzing the novel (SP) boundary
conditions.

The outline of this article is as follows: in the next section we
present the basic preliminary notions regarding the algebraic
setting for classical models on the full line and on the interval.
In our analysis we adopt the line of attack described in e.g.
\cite{ft}, and in \cite{mac, nls} for boundary systems. More
precisely we introduce the classical Yang-Baxter equation and the
underlying algebra for the system on the full line. In the situation
of a system on the interval we distinguish two types of boundary
conditions based on the classical versions of the reflection algebra
(SP) and ($q$) twisted Yangian (SNP). Next the $A_n^{(1)}$ ATFT on
the full line is reviewed and an explicit derivation of the local
integrals of motion by solving the auxiliary linear problem
\cite{ft} is presented. In section 3 being guided by the same logic
and adopting Sklyanin's formulation \cite{sklyanin} we rederive the
integrals of motion of the $A_n^{(1)}$ ATFT with SNP boundary
conditions. Note that analogous strategy was followed in \cite{mac}
and \cite{nls} for the classical boundary sine-Gordon and vector NLS
models respectively. Our results are in agreement with the ones
deduced in \cite{durham}. In section 4 we introduce for the first
time the novel boundary conditions (SP) within the context of ATFT.
Explicit expressions of the associated local integrals of motion are
deduced from the asymptotic expansion of the classical transfer
matrix. It is worth stressing that the induced integrals of motion
depend on free boundary parameters as opposed to the SNP case. In
the last section a discussion on the entailed results is presented
and several directions for future investigations are proposed.

\section{Preliminaries}

The analysis of the ATFT with integrable boundary conditions will
rely on the solution of the so called auxiliary linear problem
\cite{ft}. Before we proceed to the study of classical integrable
models with consistent boundary conditions it will be instructive
to recall the basic notions in the periodic case. Let $\Psi$ be a
solution of the following set of equations \be &&{\partial \Psi
\over
\partial x} = {\mathbb U}(x,t, \lambda)  \Psi \label{dif1}\\ &&
{\partial \Psi \over \partial t } = {\mathbb V}(x,t,\lambda)\Psi
\label{dif2} \ee with ${\mathbb U},\ {\mathbb V}$ being in general
$n \times n$ matrices with entries functions of complex valued
fields, their derivatives, and the spectral parameter $\lambda$.
Compatibility conditions of the two differential equation
(\ref{dif1}), (\ref{dif2}) lead to the zero curvature condition
\cite{AKNS, abla, ZSh} \be \dot{{\mathbb U}} - {\mathbb V}' + \Big
[{\mathbb U},\ {\mathbb V} \Big ]=0. \label{zecu} \ee The latter
equations give rise to the corresponding classical equations of
motion of the system under consideration. The monodromy matrix
from (\ref{dif1}) may be written as: \be T(x,y,\lambda) = {\cal P}
exp \Big \{ \int_{y}^x {\mathbb U}(x',t,\lambda)dx' \Big \}
\label{trans} \ee with $T(x, x , \lambda) =1$. The monodromy
matrix satisfies apparently (\ref{dif1}), and this will be
extensively used in the present analysis. On the other hand within
the Hamiltonian formalism the existence of the classical
$r$-matrix, satisfying the classical Yang-Baxter equation
\cite{skl, sts} \be \Big [r_{12}(\lambda_1-\lambda_2),\
r_{13}(\lambda_1)+r_{23}(\lambda_2) \Big]+ \Big
[r_{13}(\lambda_1),\ r_{23}(\lambda_2) \Big] =0, \ee guarantees
the integrability of the classical system. Indeed, consider the
operator $T(x,y,\lambda)$ satisfying \be \Big
\{T_{1}(x,y,t,\lambda_1),\ T_{2}(x,y,t,\lambda_2) \Big \} =
\Big[r_{12}(\lambda_1-\lambda_2),\
T_1(x,y,t,\lambda_1)T_2(x,y,t,\lambda_2) \Big ]. \label{basic} \ee
Making use of the latter equation one may readily show for a
system in full line: \be \Big \{\ln tr\{T(x,y,\lambda_1)\},\ \ln
tr\{T(x,y, \lambda_2)\} \Big\}=0 \ee i.e. the system is
integrable, and the charges in involution --local integrals of
motion-- may be obtained by expanding the object $\ln
tr\{T(x,y,\lambda)\}$.

The classical $r$-matrix associated to the $A_{n}^{(1)}$ affine
Toda field theory in particular is given by\footnote{Notice that
the $r$-matrix employed here is in fact $r_{12}^{t_1 t_2}$ with
$r_{12}$ being the matrix used e.g. in \cite{done, doikoun}}
\cite{jimbo} \be r(\lambda) = {\cosh(\lambda) \over \sinh
(\lambda)} \sum_{i=1}^{n+1} e_{ii}\otimes e_{ii} + {1\over \sinh
(\lambda)} \sum_{i \neq j =1}^{n+1} e^{ [ sgn(i-j) -(i-j) {2 \over
n+1}  ] \lambda } e_{ij} \otimes e_{ji}. \label{rc} \ee Note that
the classical $r$-matrix (\ref{rc}) is written in the so called
principal gradation as is also in \cite{durham, gand}. To express
the $r$-matrix in the homogeneous gradation one implements a
simple gauge transformation: \be r^{(h)} (\lambda) = {\cal
V}(\lambda)\ r^{(p)}(\lambda)\ {\cal V}(-\lambda) ~~~~
\mbox{where} ~~~~~{\cal V}(\lambda) = \sum_{j=1}^{n+1} e^{{2 (j-1)
\lambda \over n+1}} e_{jj}. \ee

Our main aim as mentioned upon is to study the $A_n^{(1)}$ ATFT on
the interval. For this purpose we shall employ Sklyanin's
formulation (see also \cite{mac, nls} for classical models with
integrable boundary conditions). It will be convenient for our
purposes here to introduce some useful notation: \be && \hat
r_{ab}(\lambda) = r_{ba}(\lambda) ~~~\mbox{for SP}, ~~~~\hat
r_{ab}(\lambda) = r_{ba}^{t_a t_b}(\lambda) ~~~\mbox{for SNP}
\non\\ && r^*_{ab}(\lambda) =r_{ab}(\lambda) ~~~\mbox{for SP},
~~~~ r^*_{ab}(\lambda) = r_{ba}^{t_b}(-\lambda) ~~~\mbox{for SNP}
\non\\ && \hat r^*_{ab}(\lambda) =r_{ba}(\lambda) ~~~\mbox{for
SP}, ~~~~\hat
r^*_{ab}(\lambda) = r_{ab}^{t_a}(-\lambda) ~~~\mbox{for SNP} \non\\
&&  T(\lambda) =T^{-1}(-\lambda) ~~~\mbox{for SP}, ~~~~\hat
T(\lambda) = T^t(-\lambda) ~~~\mbox{for SNP}. \label{notation00}
\ee In the situation where non-trivial integrable boundary
conditions are implemented one derives two types of `monodromy'
matrices ${\cal T}$, which respectively represent the classical
versions of the reflection algebra ${\mathbb R}$, and the twisted
Yangian ${\mathbb T}$ written in the compact form below (see e.g.
\cite{sklyanin, maillet}): \be && \Big \{{\cal T}_1(\lambda_1),\
{\cal T}_2(\lambda_2) \Big \} = r_{12}(\lambda_1-\lambda_2){\cal
T}_{1}(\lambda_1){\cal T}_2(\lambda_2) -{\cal T}_1(\lambda_1)
{\cal T}_2(\lambda_2) \hat r_{12}(\lambda_1 -\lambda_2) \non\\ &&
+ {\cal T}_{1}(\lambda_1) \hat r^*_{12}(\lambda_1+\lambda_2){\cal
T}_2(\lambda_2)- {\cal T}_{2}(\lambda_2)
r^*_{12}(\lambda_1+\lambda_2){\cal T}_1(\lambda_1). \label{refc}
\ee The modified `monodromy' matrices, compatible with the
corresponding algebras ${\mathbb R},\ {\mathbb T}$ are given by
the following expressions \cite{sklyanin, durham}: \be && {\cal
T}(x,y,t,\lambda) = T(x,y,t,\lambda)\ K^{-}(\lambda)\ \hat
T(x,y,t,\lambda) \label{reps} \ee and the generating function of
the involutive quantities is defined as \be t(x,y,t,\lambda)= tr\{
K^{+}(\lambda)\ {\cal T}(x,y,t,\lambda)\} \ee  where $K^{\pm}$
$c$-number representations of the algebra ${\mathbb R}$ (${\mathbb
T}$) satisfying (\ref{refc}) for SP and SNP respectively, and also
\be \Big \{ K_1^{\pm}(\lambda_1),\ K_2^{\pm}(\lambda_2) \Big \}=0.
\ee Due to (\ref{refc}) it can be shown that \be  \Big
\{t(x,y,t,\lambda_1),\ t(x,y,t,\lambda_2) \Big \} =0, ~~~
\lambda_1,\ \lambda_2 \in {\mathbb C}. \label{bint} \ee Technical
details on the proof of classical integrability are provided e.g.
in \cite{sklyanin, durham, nls}.

\subsection{Classical integrals of motion in ATFT}

We shall exemplify our investigation using the first non-trivial
model of the ATFT hierarchy that exhibits both types of boundary
conditions, that is the $A_2^{(1)}$ case. Recall the Lax pair for
a generic $A_{n}^{(1)}$ theory \cite{olive}: \be && {\mathbb
V}(x,t,u) = {\beta \over 2}\
\partial_x \Phi \cdot H + {m \over 4}\ \Big (u\ e^{{\beta \over 2}
\Phi \cdot H}\ E_{+}\ e^{- {\beta \over 2} \Phi \cdot H} -{1\over
u}\ e^{-{\beta \over 2}
\Phi \cdot H}\ E_{-}\ e^{{\beta \over 2} \Phi \cdot H}\Big) \non\\
&& {\mathbb U}(x,t,u) = {\beta \over 2}\ \Pi \cdot H + {m \over
4}\ \Big (u\ e^{{\beta \over 2} \Phi \cdot H}\ E_{+}\ e^{- {\beta
\over 2} \Phi \cdot H} +{1\over u}\ e^{-{\beta \over 2} \Phi \cdot
H}\ E_{-}\ e^{{\beta \over 2} \Phi \cdot H}\Big)  \label{lpair}
\ee $\Phi,\ \Pi$ are $n$-vector fields, with components $\phi_i,\
\pi_i,\ i \in \{1, \ldots , n\}$, $~u=e^{{2\lambda \over n+1}}$ is
the multiplicative spectral parameter. To compare with the
notation used in \cite{durham} we set ${m^2 \over 16}= {\tilde m^2
\over 8}$ ($\tilde m$ denotes the mass in \cite{durham}). Note
that eventually in \cite{durham} both $\beta,\ \tilde m$ are set
equal to unit.

Also define: \be E_+ = \sum_{i=1}^{n+1} E_{\alpha_i},
~~~~~E_{-}=\sum_{i=1}^{n+1} E_{-\alpha_i} \ee $\alpha_i$ are the
simple roots, $H$ ($n$-vector) and $E_{\pm \alpha_i}$ are the
algebra generators in the Cartan-Weyl basis corresponding to
simple roots, and they satisfy the Lie algebra relations: \be &&
\Big [H,\ E_{\pm \alpha_i} \Big]= \pm \alpha_i E_{\pm \alpha_i},
\non\\ && \Big [E_{\alpha_i},\ E_{-\alpha_i} \Big ] = {2\over
\alpha_i^2}\ \alpha_i \cdot H \ee Explicit expressions on the
simple roots and the Cartan generators are presented in Appendix
A. Notice that the Lax pair has the following behavior: \be
{\mathbb V}^t(x, t, -u^{-1}) = {\mathbb V}(x,t, u), ~~~~~{\mathbb
U}^t(x, t, u^{-1}) = {\mathbb U}(x,t,u) \ee where $^t$ denotes
usual transposition.

Our objective as mentioned is to examine the system with non-trivial
boundaries, thus we consider representations of the associated
underlying algebras expressed by ${\cal T}$. To recover the local
integrals of motion of the considered system we shall follow the
quite standard procedure and expand $\ln\ t(u)$ in powers of
$u^{-1}$. An alternative strategy would be to derive the modified
Lax pair, compatible with the boundary conditions chosen, and hence
the associated equations of motion (see e.g. \cite{durham}). A
systematic derivation of boundary Lax pairs independently of the
choice of model is discussed in \cite{avandoikou}. To expand the
open transfer matrix and derive the local integrals of motion we
shall need the expansions of $T(x,y, u)$, $~T(x,y, u^{-1})$ and
$K^{\pm}(u)$. In what follows in the present section we basically
introduce the necessary preliminaries for such a derivation, and we
also reproduce the known integrals of motion for the ATFT on the
full line.

Let $T'(x, y, u) = T(x, y, u^{-1})$ and ${\mathbb U}'(x, u ) =
{\mathbb U}(x, u^{-1})$. Following the logic described in
\cite{ft} for the sine-Gordon model, we aim at expressing the part
associated to $E_+$, $E_-$ in ${\mathbb U},\ {\mathbb U}'$
respectively independently of the fields, after applying a
suitable gauge transformation. More precisely, consider the
following gauge transformation: \be &&
T(x, y, u )= \Omega(x)\ \tilde T(x,y, u)\ \Omega^{-1}(y), \non\\
&& T'(x, y ,u) = \Omega^{-1}(x)\ \tilde T'(x,y,u)\ \Omega(y)
~~~~~~~~\Omega(x) = e^{{\beta \over 2} \Phi(x)\cdot H}. \ee Then
from  equation (\ref{dif1}) we obtain that the gauge transformed
operators ${\mathbb U},\ {\mathbb U}'$ can be expressed as: \be &&
\tilde {\mathbb U}(x,t,u) = \Omega^{-1}(x)\ {\mathbb
U}(x,t,u)\ \Omega(x) - \Omega^{-1}(x)\ {d \Omega(x)\over dx} \non\\
&& \tilde {\mathbb U}'(x,t,u) = \Omega(x)\ {\mathbb U}'(x,t,u)\
\Omega^{-1}(x) - \Omega(x)\ {d \Omega^{-1}(x)\over dx}. \ee After
implementing the gauge transformations the operators $\tilde
{\mathbb U},\ \tilde {\mathbb U}'$ take the following simple form:
\be \tilde {\mathbb U}(x,t, u) = {\beta \over 2} \Theta \cdot H
+{m\over 4} \Big ( u E_+ + {1 \over u} X_- \Big ), ~~~~~ \tilde
{\mathbb U}' (x, t, u) = {\beta \over 2} \hat \Theta \cdot H
+{m\over 4} \Big ( u E_- + {1 \over u} X_+ \Big ) \ee where we
define: \be \Theta = \Pi -
\partial_x \Phi,~~~~~\hat \Theta = \Pi + \partial_x \Phi, ~~~~~X_- =
e^{- \beta \Phi \cdot H}\ E_-\ e^{\beta  \Phi \cdot H}, ~~~~~X_+ =
e^{ \beta \Phi \cdot H}\ E_+\ e^{-\beta  \Phi \cdot H} \ee $\tilde
T,\ \tilde {\mathbb U}$ also satisfy (\ref{dif1}), and $\Theta,\
\hat \Theta$ are $n$ vectors with components $\theta_i,\ \hat
\theta_i$ respectively.

Consider now the following ansatz for $\tilde T$, $~\tilde T'$ as
$|u| \to \infty$ \cite{ft} \be && \tilde T(x,y,u) = ({\mathbb I}
+W(x, u))\ \exp[Z(x,y,u)]\ ({\mathbb I} +W(y,u))^{-1}, \non\\
&&\tilde T'(x,y, u) = ({\mathbb I}+ \hat W(x, u))\ \exp[\hat
Z(x,y,u)]\ ({\mathbb I} + \hat W(y,u))^{-1}, \label{exp0} \ee
where $W,\ \hat W$ are off diagonal matrices i.e. $~W =
\sum_{i\neq j} W_{ij} E_{ij}$, and $Z,\ \hat Z$ are purely
diagonal $~Z = \sum_{i=1}^{n+1} Z_{ii}E_{ii}$. Also \be Z(u) =
\sum_{k=-1}^{\infty} {Z^{(k)} \over u^{k}}, ~~~~W_{ij} =
\sum_{k=0}^{\infty}{W^{(k)} \over u^k}. \label{expa} \ee Inserting
the latter expressions (\ref{expa}) in (\ref{dif1}) one may
identify the coefficients $W_{ij}^{(k)}$ and $Z_{ii}^{(k)}$.
Indeed from (\ref{dif1}) we obtain the following fundamental
relations: \be && {d Z \over d x} = {\tilde \mathbb U}^{(D)}  +
({\tilde \mathbb U}^{(O)}\ W)^{(D)} \non\\
&& {d  W\over dx } + W {\tilde \mathbb U}^{(D)} -{\tilde \mathbb
U}^{(D)}W + W({\tilde \mathbb U}^{(O)}W)^{(D)} -{\tilde \mathbb
U}^{(O)} - ({\tilde \mathbb U}^{(O)}W)^{(O)}=0 \label{form} \ee
where the superscripts $O,\ D$ denote off-diagonal and diagonal
part respectively. Similar relations may be obtained for $\hat Z,\
\hat W$, in this case $\tilde {\mathbb U} \to \tilde {\mathbb
U}'$. We omit writing these equations here for brevity.

It will be useful in what follows to introduce some notation: \be
{\beta \over 2} \Theta \cdot H = \mbox{diag} (a,\ b,\ c),
~~~~~{\beta \over 2} \hat \Theta \cdot H = \mbox{diag} (\hat a,\
\hat b,\ \hat c), ~~~~ e^{\beta \alpha_i \cdot \Phi} = \gamma_i
\label{def} \ee explicit expression of $a,\ b,\ c$ and $\gamma_i$
can be found in Appendix B (\ref{def2}); notice that $a +b +c =0$.
From the first of equations (\ref{form}) we may derive the
matrices $Z,\ \hat Z$. Indeed one may easily show that: \be {d
Z^{(0)}\over dx } &=& {m\over 4} \left(
\begin{array}{ccc}
 W_{21}^{(1)} + \zeta a     &  & \\
                    & W_{32}^{(1)} + \zeta b &  \\
 &  &  -W_{13}^{(1)} + \zeta c\\ \end{array} \right) =0 \non\\
{d \hat Z^{(0)}\over dx }   &=& {m\over 4} \left(
\begin{array}{ccc}
-\hat W_{31}^{(1)}  +\zeta \hat a    &  & \\
                                          & \hat W_{12}^{(1)}  + \zeta \hat b &  \\
 &  & \hat W_{23}^{(1)}  +\zeta \hat c\\ \end{array} \right)=0 \ee
it is clear that the latter quantities are zero because of the
form of $W_{ij}^{(1)},\ \hat W_{ij}^{(1)}$ see Appendix B. Also
the higher order $Z^{(k)},\ \hat Z^{(k)}$ are given by: \be {d
Z^{(k)}\over dx }  &=& {m\over 4} \left(
\begin{array}{ccc}
W_{21}^{(k+1)}  -\gamma_3 W_{31}^{(k-1)}     &  & \\
                                          & W_{32}^{(k+1)}  +\gamma_1 W_{12}^{(k-1)} &  \\
 &  & -W_{13}^{(k+1)}  +\gamma_2 W_{23}^{(k-1)}\\ \end{array} \right) \non\\
{d \hat Z^{(k)}\over dx }   &=& {m\over 4} \left(
\begin{array}{ccc}
-\hat W_{31}^{(k+1)}  +\gamma_1 \hat W_{21}^{(k-1)}     &  & \\
                                          & \hat W_{12}^{(k+1)}  +\gamma_2 \hat W_{32}^{(k-1)} &  \\
 &  & \hat W_{23}^{(k+1)}  -\gamma_3 \hat  W_{13}^{(k-1)}\\ \end{array} \right)
\non\\ k > 0. \label{zz0} \ee The computation of $W,\ \hat W$ is
essential for defining the diagonal elements. First it is
important to discuss the leading contribution of the above
quantities as $|u| \to \infty$. To achieve this we shall need the
explicit form of $Z^{(-1)},\ \hat Z^{(-1)}$: \be Z^{(-1)}(x, y) =
{m (x-y)\over 4} \left( \begin{array}{ccc}
e^{{i \pi \over 3}}     &  & \\
                                          &e^{-{i \pi \over 3}}  &  \\
 &  & -1 \\ \end{array} \right), ~~~\hat Z^{(-1)}(x,y)=  {m (x-y)\over 4}
\left( \begin{array}{ccc}
e^{-{i \pi \over 3}}     &  & \\
                                          &e^{{i \pi \over 3}}  &  \\
 &  & -1 \\ \end{array} \right). \non\\ \label{zz}
\ee The information above will be extensively used in what
follows.

Before we proceed with the analysis of integrable boundary
conditions in ATFT let us first reproduce the known local
integrals of motion in the periodic case, emerging from the
expansion ($|u|\to \infty$) \be \ln\ [tr T(u)] = \ln\ [tr
\{(1+W(L,u))\ e^{Z(L, -L, u)}\ (1+W(-L,u))^{-1} \}]. \ee Notice
that in the case of periodic boundary conditions we put our system
in the `whole' line ($x = L,\ y =-L$), and consider Schwartz
boundary conditions, i.e. the fields and their derivatives vanish
at the end points $\pm L$. Bearing in mind that as $u \to -\infty$
the leading contribution of $e^{Z},\ (e^{\hat Z})$ (see
(\ref{zz})) comes from the $e^{Z_{33}},\ (e^{\hat Z_{33}})$ term,
the expression above becomes \be \ln\ [tr T(u \to -\infty)] =
\sum_{k= -1} {Z^{(k)}_{33} \over u^k}. \ee To reproduce the
familiar local integrals of motion we shall need both $Z(L, -L,
u),\ \hat Z(L, -L, u)$. Let \be &&{\cal I}_1 = -{12 m \over
\beta^{2} } Z_{33}^{(1)}(L, -L, u) = \int_{-L}^{L} dx \Big(
\sum_{i=1}^2 \theta_i^2  + {m^2 \over \beta^2}
\sum_{i=1}^3 e^{\beta \alpha_i \cdot \Phi}  \Big ), \non\\
&& {\cal I}_{-1} = -{12 m \over \beta^{2} }
\hat Z_{33}^{(1)}(L, -L, u) = \int_{-L}^{L} dx
\Big( \sum_{i=1}^2 \hat \theta_i^2 + {m^2 \over \beta^2}
\sum_{i=1}^3 e^{\beta \alpha_i \cdot \Phi}  \Big ) \non\\
&&{\cal I}_2 = {3 m^2 \over 2 \beta^3} Z_{33}^{(2)} (L, -L, u) =
\int _{-L}^L dx \Big ( {8 \over \beta^3}(a b c - b c') -  {m^2 \over 2 \beta ^3}
(\gamma_1 c +\gamma_2 a+ \gamma_3 b )\Big ) \non\\
&&{\cal I}_{-2} = {3 m^2 \over 2 \beta^3}  \hat Z_{33}^{(2)} (L,
-L, u) = \int _{-L}^L dx \Big ( {8 \over \beta^3}(\hat a \hat b
\hat c + \hat b \hat c') - {m^2 \over 2 \beta ^3} (\gamma_1 \hat c
+\gamma_2 \hat a+ \gamma_3 \hat b )\Big ) \non\\ && \ldots
~~~(\mbox{higher local integrals of motion}) \ee the momentum and
Hamiltonian (and the higher conserved quantities) of the ATFT are
given by: \be && {\cal P}_{1} = {1\over 2}({\cal I}_{-1} - {\cal
I}_{1} ) = \int_{-L}^{L} dx
\sum_{i=1}^2 \Big ( \pi_i\ \phi_i' - \pi_i'\ \phi_i \Big ) \non\\
&& {\cal H}_{1} = {1\over 2} ({\cal I}_1 +{\cal I}_{-1}) =
\int_{-L}^{L} dx \Big( \sum_{i=1}^2 (\pi_i^2 + \phi_i^{'2}) + {m^2
\over \beta^2}
\sum_{i=1}^3 e^{\beta \alpha_i \cdot \Phi}  \Big ) \non\\
&& {\cal P}_{2} = {1 \over 2}({\cal I}_{-2} - {\cal I}_{2} ) \non\\
&&={1 \over 2} \int_{-L}^L dx
\Big ({8\over \beta^3}( \hat a \hat b \hat c- abc ) +{8\over \beta^3}
(bc' +\hat b \hat c') +{m^2 \over 2\beta^3}(\gamma_1(c-\hat c) +\gamma_2(a-\hat a)
+\gamma_3 (b-\hat b)) \Big ) \non\\
&& {\cal H}_{2} = {1\over 2} ({\cal I}_2 +{\cal I}_{-2}) \non\\
&&= {1 \over 2} \int_{-L}^L dx \Big ({8\over \beta^3}( abc +\hat a
\hat b \hat c) -{8\over \beta^3} (bc' -\hat b \hat c') -{m^2 \over
2\beta^3}(\gamma_1(c+\hat c) +\gamma_2(a+\hat a) +\gamma_3 (b+\hat
b)) \Big ) \non\\ && \ldots  \label{im} \ee Note that the boundary
terms are absent in the expressions above, since we considered
Schwartz type boundary conditions. Also, in the generic situation,
for any $A_{n}^{(1)}$, the sum in the momentum ${\cal P}_1$ and the
kinetic term of the Hamiltonian ${\cal H}_1$ runs from 1 to $n$,
whereas the sum in the potential term of the Hamiltonian runs from 1
to $n+1$.

\section{SNP boundary conditions}

We turn now to our main concern, which is the study of integrable
boundary conditions in ATFT. We shall first discuss the boundary
conditions that already have been analyzed in \cite{durham}. Based
on the underlying algebra, that is the classical analogue of the
$q$-twisted Yangian we shall reproduce the previously known
results \cite{durham}, so this section serves basically as a warm
up exercise. In the subsequent section we shall analyze in detail
the novel boundary conditions (SP) associated to the classical
version of the reflection algebra.

To obtain the relevant local integrals of motion we shall expand
the following object (consider now $x=0,\ y=-L$): \be
\ln\ t(u) &=& \ln\ tr \Big \{ K^+(u)\ T(u)\ K^-(u)\ T^t(u^{-1}) \Big \}\non\\
&=& \ln\ tr \Big \{ K^+(u)\ \Omega(0)\ \tilde T(u)\
\Omega^{-1}(-L)\ K^-(u)\ \Omega(-L)\ \tilde T^t(u^{-1})\
\Omega^{-1}(0)\Big \} \ee For simplicity here, but without really
losing generality we consider Schwartz boundary conditions at the
boundary point $-L$ and $K^-(u) \propto {\mathbb I}$. Also $K^+(u)
= K^t(u^{-1})$ where $K$ is any $c$-number solution of the twisted
Yangian. Taking also into account the ansatz (\ref{exp0}) we
conclude \be && \ln\ t(u) = \ln\ tr \Big \{(1 +\hat W^t(0, u))
\Omega^{-1}(0)\ K^+(u)\ \Omega (0)\ (1 +W(0, u))\ e^{Z(0, -L, u)+
\hat Z(0, -L, u)}\Big \}. \non\\ \ee Recall from the previous
section that as $u \to -\infty$ the leading contribution of
$e^{Z},\ e^{\hat Z}$ comes from the $e^{Z_{33}},\ e^{\hat Z_{33}}$
terms (see (\ref{zz})), hence \be \ln\ t(u) &=& Z_{33}(0, -L, u) +
\hat Z_{33} (0, -L, u) + \ln [ (1 +\hat W^t(0, u)) \Omega^{-1}(0)\
K^+(u)\ \Omega (0)\
(1 +W(0, u))]_{33} \non\\
&=& \sum_{k=-1}^{\infty} {Z_{33}^{(k)} +\hat Z_{33}^{(k)}  \over
u^k} + \sum_{k=0}^{\infty} {{\mathrm f}_k \over u ^k}.
\label{expan2}  \non\\
\ee To obtain the explicit form of the boundary contributions to
the integrals of motion we should first review known results on
the solution of the reflection equation for SNP boundary
conditions. The generic solution for the $A_{n}^{(1)}$ case in the
principal gradation are given by \cite{gand, ann2}: \be &&
K(\lambda) = (g e^{\lambda} + \bar g e^{-\lambda})
\sum_{i=1}^{n+1} e_{ii} + \sum_{i>j} f_{ij} e^{ \lambda -
{2\lambda \over n+1} (i-j)} e_{ij}+ \sum_{i<j} f_{ij} e^{ -\lambda
- {2\lambda \over n+1} (i-j)} e_{ij} \non\\ && g =q^{-{1\over 2} +
{n+1 \over 4}} ~~~~\bar g = \pm q^{{1\over 2} - {n+1 \over 4}},
~~~f_{ij} = \pm q^{-{n+1 \over 4}}, ~~~~f_{ji} = q^{{n+1 \over
4}}, ~~~i < j. \label{kp} \ee In order to effectively compare with
the results of \cite{gand} as well as being compatible with
\cite{durham} we always express in the text both $r$ and $K$
matrices in the principal gradation. Nevertheless, to obtain the
matrix in the homogenous gradation as given in \cite{ann2} we
implement the following gauge transformation \be K^{(h)} = {\cal
V}(\lambda)\ K^{(p)}(\lambda)\ {\cal V}(-\lambda). \ee

We shall now focus on the $A_2^{(1)}$ case, which is our main
example here. Recall that $K^+(u) = K^t(u^{-1})$ then the
$K^+$-matrix is $3 \times 3$ matrix written explicitly as: \be &&
K^+(u) = u^{{3\over 2}} \bar G + u^{{1\over 2}} \bar F +
u^{-{1\over 2}} F + u^{-{3\over 2}} G ~~~~~\mbox{where} \non\\
&& G = g\ {\mathbb I}, ~~~\bar G = \bar g\ {\mathbb I}, \non\\
&& \bar F = f_{12}\ e_{21} + f_{23}\ e_{32} + f_{31}\ e_{13}, \non\\
&& F = f_{21}\ e_{12} +f_{32}\ e_{23}  +f_{13}\ e_{31} \label{k1}
\ee and the coefficients $g,\ \bar g,\ f_{ij}$ are given in
(\ref{kp}) with $n=2$. Bearing in mind the explicit form of the
boundary matrix we may identify the factors ${\mathrm f}_i$ in the
expansion (\ref{expan2}) which are reported in Appendix C. Taken
into account expressions (\ref{expan2}), (\ref{boundary}) and
$Z^{(1)}_{33},\ \hat Z^{(1)}_{33}$ given in Appendix B we conclude
for the first non-trivial boundary integral of motion: \be {\cal
H}_1^{(b)} &=& -{6m \over \beta^2}\Big ( Z_{33}^{(1)} +\hat
Z_{33}^{(1)} +{\mathrm f}_1\Big ) \non\\ &=& \int_{-L}^{0} dx
\Big( \sum_{i=1}^2 (\pi_i^2 + \phi_i^{'2}) + {m^2 \over \beta^2}
\sum_{i=1}^3 e^{\beta \alpha_i \cdot \Phi}  \Big ) +{2 m \over
\bar g \beta^2} \Big ( f_{12} e^{{\beta \over 2} \alpha_1 \cdot
\Phi(0)} +f_{23} e^{{\beta \over 2} \alpha_2 \cdot \Phi(0)}
-f_{31}
e^{{\beta \over 2} \alpha_3 \cdot \Phi(0)}   \Big ). \non\\
\label{bham} \ee In general for the $A_{n}^{(1)}$ ATFT the
boundary Hamiltonian with SNP boundary conditions will have the
following from \be {\cal H}_1^{(b)} = \int_{-L}^{0} dx \Big(
\sum_{i=1}^n (\pi_i^2 + \phi_i^{'2}) + {m^2 \over \beta^2}
\sum_{i=1}^{n+1} e^{\beta \alpha_i \cdot \Phi}  \Big ) +
\sum_{i=1}^{n+1} c_i\ e^{{\beta \over 2} \alpha_i \cdot \Phi(0)},
\ee which as expected coincides with the boundary Hamiltonian
deduced in \cite{durham}. It is quite easy to check that in the
case of a trivial boundary conditions, i.e. $K^+ \propto {\mathbb
I}$ the boundary terms containing $c_i$ disappear and the entailed
Hamiltonian has exactly the same structure as in the bulk case.

The second conserved charge of the hierarchy is given by \be &&
{\cal H}^{(b)}_{2} = {3 m^2 \over 4 \beta^3 } (Z_{33}^{(2)}
+ \hat Z_{33}^{(2)} + {\mathrm f}_{2}) \non\\
&&= {1 \over 2} \int_{-L}^0 dx \Big ({8\over \beta^3}( abc +\hat a
\hat b \hat c) -{8\over \beta^3} (bc' -\hat b \hat c') -{m^2 \over
2\beta^3}(\gamma_1(c+\hat c) +\gamma_2(a+\hat a) +\gamma_3 (b+\hat
b)) \Big ) \non\\
&& -{m^2 \over 4 \bar g \beta^2} \Big ( f_{21} e^{-{\beta \over 2}
\alpha_1 \cdot \Phi(0)} + f_{32} e^{-{\beta \over 2} \alpha_2 \cdot
\Phi(0)} -f_{13} e^{-{\beta \over 2} \alpha_3 \cdot \Phi(0)} \Big )
+ {4 \over \beta^2} \Big ( \hat c^2(0) -\hat a(0) c(0) -b(0) \hat
c(0) \Big )
\non\\
&& - {m \over \bar g \beta^2} \Big ( f_{12} (c(0) +\hat c(0))
e^{{\beta \over 2} \alpha_1 \cdot \Phi(0)}  -f_{23} b(0) e^{{\beta
\over 2} \alpha_2 \cdot \Phi(0)} + f_{31} \hat a(0)  e^{{\beta
\over 2} \alpha_3 \cdot \Phi(0)} \Big ) \non\\ && -{3 m^2 \over 8
\bar g \beta^2} \Big (-{1\over 3 \bar g}(f_{12} e^{{\beta \over 2}
\alpha_1 \cdot \Phi(0)} + f_{23} e^{{\beta \over 2} \alpha_2 \cdot
\Phi(0)} -f_{31} e^{{\beta \over 2} \alpha_3 \cdot \Phi(0)} ) +
{\zeta \over 3}(c(0)- b(0) +\hat c(0) - \hat a(0)) \Big )^2.
\non\\ \label{bham2} \ee Again when assuming the simplest boundary
conditions $K^+ \propto {\mathbb I}$ we conclude that all the
boundary terms containing the factors $f_{ij}$ disappear, exactly
as it happens in the first Hamiltonian. The bulk parts of the
boundary Hamiltonians above coincide with the ones found in the
previous section --for Schwartz type boundary conditions. Extra
boundary terms are added due to the presence of the non-trivial
$K$-matrix.

Notice that the boundary analogues of ${\cal P}_k$ are not
conserved quantities anymore similarly to the sine-Gordon model on
the half line, where only the `half' of the bulk charges are
conserved after the implementation of consistent integrable
boundary conditions. We should stress that this is a consequence
of the particular choice of boundary conditions, and this will
become apparent in the next section while analyzing the novel
boundary conditions. Note also that in the expressions for the
boundary Hamiltonian written above there exist no free boundary
parameter, contrary to the SP case as will see subsequently.
Analogous results may be seen in the context of quantum integrable
spin chains regarding the explicit expression of the corresponding Hamiltonians
as well as their symmetries \cite{doikousnp,
ann1, ann2}.

Let us finally mention that one can in general consider `dynamical'
boundary conditions (see e.g. \cite{baz, bade, nls}). In this case
instead of assuming a $c$-number solution of the classical version
of the $q$-twisted Yangian (\ref{refc}) we consider a generic
--dynamical-- representation of the algebra defined as
\cite{sklyanin}: \be {\mathbb K}(\lambda) = L(\lambda -\Theta)\
K(\lambda) \otimes {\mathbb I}\ L^{t}(-\lambda -\Theta) \ee where
$K$ is a $c$-number solution of the classical twisted Yangian
\cite{gand, ann2}, and $L$ is any solution of the fundamental
relation (\ref{basic}) e.g. a $q$-oscillator. Such boundary
conditions for the $A_2^{(1)}$ ATFT have been analyzed in
\cite{baz}. More precisely, in this case the entries of ${\mathbb
K}$ are not $c$-number anymore, but algebraic objects satisfying
Poisson commutation relations dictated by the underlying classical
algebra. At the quantum level these objects, and consequently the
quantities $f_{ij}$ appearing in the local integrals of motion
(\ref{bham}), (\ref{bham2}), become operators obeying commutation
relation defined by the $q$-twisted Yangian. In fact, due to the
`dynamical nature' of the boundary conditions extra degrees of
freedom, incorporated in $L$, are attached to the boundary.

\section{SP boundary conditions}

We come now to the study of the more intriguing, at least in the
present context, boundary conditions. Here for the first time we
systematically analyze the new boundary conditions (SP) starting
from the underlying algebra i.e. the reflection algebra. In this
case the generating function of the local integrals of motion is
given by the following expression: \be
\ln\ t(u) &=& \ln\ tr \Big \{ K^+(u)\ T(u)\ K^-(u)\ T^{-1}(u^{-1}) \Big \}\non\\
&=& \ln\ tr \Big \{ K^+(u)\ \Omega(0)\ \tilde T(u)\
\Omega^{-1}(-L)\ K^-(u)\ \Omega^{-1}(-L)\ \tilde T^{-1}(u^{-1})\
\Omega(0)\Big \} \ee taking into account the ansatz (\ref{exp0})
we conclude \be  \ln\ t(u) &=& \ln\ tr \Big \{(1 +\hat W(0,
u))^{-1} \Omega(0)\ K^+(u)\ \Omega (0)\ (1 +W(0, u))\ e^{Z(0, -L,
u)}\non\\ && (1+W(-L,u))^{-1} \Omega^{-1}(-L) K^-(u)
\Omega^{-1}(-L)  (1 +\hat W(-L, u)) e^{-\hat Z(0, -L, u)}\Big \}.
\ee The leading contribution of $e^{Z},\ e^{-\hat Z}$ comes from
the $e^{Z_{11}},\ e^{-\hat Z_{11}}$ terms as $i u \to \infty$,
whereas as $i u \to -\infty$ it comes from the $e^{Z_{22}},\
e^{-\hat Z_{22}}$ terms. Depending on the limit we assume we
obtain two distinct expressions for $i u \to \infty$ and $i u \to
-\infty$ respectively: \be \ln\ t(i u \to \infty) &=& Z_{11}(0,
-L, u) - \hat Z_{11} (0, -L, u) \non\\ &+& \ln [ (1 +\hat W(0,
u))^{-1} \Omega(0)\ K^+(u)\ \Omega (0)\
(1 +W(0, u))]_{11} \non\\
&+& \ln [ (1
+ W(-L, u))^{-1} \Omega^{-1}(-L)\ K^-(u)\ \Omega^{-1} (-L)\
(1 + \hat W(-L, u))]_{11}
\label{expan3}  \non\\
\ln\
t(iu \to -\infty) &=&  Z_{22}(0, -L, u) - \hat Z_{22} (0, -L, u) \non\\ &+& \ln [ (1
+\hat W(0, u))^{-1} \Omega(0)\ K^+(u)\ \Omega (0)\
(1 +W(0, u))]_{22} \non\\
&+& \ln [ (1 + W(-L, u))^{-1} \Omega^{-1}(-L)\ K^-(u)\ \Omega^{-1}
(-L)\ (1 + \hat W(-L, u))]_{22}. \non\\  \label{bound2} \ee
Expanding all the terms above we get \be \ln\ t(i u \to \infty)
&=& \sum_{k=-1}^{\infty} {Z_{11}^{(k)} -\hat Z_{11}^{(k)}  \over
u^n} + \sum_{k=0}^{\infty} {{\mathrm f}^+_k  +{\mathrm f}_k^-\over u^k} \non\\
\ln\ t(i u \to -\infty)  &=& \sum_{k=-1}^{\infty} {Z_{22}^{(k)}
-\hat Z_{22}^{(k)}  \over u^k} + \sum_{k=0}^{\infty} {{\mathrm
h}^+_k  +{\mathrm h}_k^-\over u ^k}. \label{expp2} \ee Although we
follow exactly the same analysis as in the SNP case, we see that
the investigation of the SP boundary conditions is technically
more involved mainly due to the fact that one has to consider the
behavior of the transfer matrix for both $i u \to \infty$ and $-i
u \to \infty$. Another technically intriguing point is that the
behavior of $(1 +W)^{-1}$, which is quite intricate, must be
studied even if the system is considered on the half line i.e.
Schwartz type boundary conditions are set at the boundary point
$-L$ (see for instance the previous section).

We shall focus here for simplicity only on diagonal solutions of the
reflection equation \cite{dvg} given by the following expressions
(in the principal gradation): \be K_{(l)}(\lambda,\ \xi) = \sinh
(\lambda + i\xi) e ^{-\lambda} \sum_{j=1}^l e^{-{4\lambda \over
n+1}(j-1)}e_{jj} + \sinh (-\lambda + i \xi)e ^{\lambda}
\sum_{j=l+1}^n e^{-{4\lambda \over n+1}(j-1)}e_{jj} \ee (recall $u =
e^{2\lambda \over n+1}$ ). To obtain the $K$-matrix in the
homogeneous gradation we implement a gauge transformation: \be
K_{(l)}^{(h)}(\lambda,\ \xi) = {\cal V}(\lambda)\
K_{(l)}^{(p)}(\lambda,\ \xi)\ {\cal V}(\lambda).\ee In fact, the
presence of non-diagonal boundary conditions does not modify the
structure of the local integrals of motion, but simply gives rise to
more complicated boundary terms.

Note that in the $A_2^{(1)}$ case we end up with two types of
diagonal boundary matrices corresponding to the two possible
values $l=1,\ 2$. We shall consider an example here to demonstrate
how the particular choice of boundary $K$-matrix contributes to
the integrals of motion. Specifically, to obtain the most general
results with the least effort it is practical to consider a
non-trivial left boundary described by $K_{(1)}$, and a right
boundary described by the $K_{(2)}$-matrix i.e. \be K^{+}(u,\
\xi^+)= K_{(1)}(u^{-1},\ \xi^+), ~~~~~~K^-(u,\ \xi^-) =K_{(2)}(u,\
\xi^-) \ee The integrals of motion emerging from the first order
of the asymptotics of the transfer matrix as $i u \to \pm \infty$
are given by: \be {\mathbb I}_1 &= & Z_{11}^{(1)} - \hat Z_{11}^{(1)}
+ {\mathrm f}_1^+ + {\mathrm f}_1^- = -{\beta^2 \over 12m} ( {\cal P}^{(b)}_1 + i \sqrt 3{\cal H}^{(b)}_1) ,
\non\\ \tilde {\mathbb I}_1 &=&
Z_{22}^{(1)} - \hat Z_{22}^{(1)} + {\mathrm h}_1^+ + {\mathrm
h}_1^- =  -{\beta^2 \over 12m} ( {\cal P}^{(b)}_1 - i \sqrt 3{\cal H}^{(b)}_1) \label{zz2} \ee (expressions for $Z,\ \hat Z, ~{\mathrm f}^{\pm}_i, ~{\mathrm h}_i^{\pm}$ are provided
in Appendix B). The momentum and energy are directly obtained from
the above conserved quantities and defined as: \be  {\cal P}_1^{(b)} &=&
\int_{-L}^0 dx \sum_{i=1}^2 \Big (\pi_i\ \phi_i' - \pi_i'\ \phi_i
\Big ) + \sum_{i=1}^2 \pi_i(0)\ \phi_i(0) + {8 \over \beta}
\alpha_2 \cdot \Pi(0)  + {12  m \over \beta^2} e^{-2i \xi^+}
e^{-\beta \alpha_3 \cdot \Phi(0)} \non\\ & -& \sum_{i=1}^2
\pi_i(-L)\ \phi_i(-L) - {8 \over \beta}  \alpha_1 \cdot \Pi(-L) +
{12 m
\over \beta^2} e^{-2i \xi^-} e^{-\beta \alpha_3 \cdot \Phi(-L)} \non\\
{\cal H}_1^{(b)} &=& \int_{-L}^0 dx \Big ( \sum_{i=1}^2
(\pi_i^2 + \phi_i^{'2} ) + {m^2 \over \beta^2} \sum_{i=1}^3 e
^{\beta \alpha_i \cdot \Phi} \Big ) + {8 \over \beta } \alpha_2
\cdot \Phi'(0)  - {8 \over \beta } \alpha_1 \cdot \Phi'(-L).
\label{ph} \ee Notice the presence of the free boundary parameters
$\xi^{\pm}$\footnote{The parameters $\xi^+,\ \xi^-$ are associated
to the right left boundary respectively. Note also that there is an
implicit dependence on the integers $l^{\pm}$.} in the local
integrals of motion above, as opposed to the SNP case where no free
boundary parameters appear in the corresponding integrals of motion.
Naturally the two boundary cases are qualitatively distinguished; in
SNP the $c$-number $K$-matrix contains no free parameters, and
consequently no free parameters occur in the entailed integrals of
motion. In the SP case however the $K$-matrix contains free
parameters, which explicitly appear in the boundary integrals of
motion. The implementation of non-diagonal $K$-matrices would lead
to the appearance of extra boundary terms and parameters in the
induced local integrals of motion.

The integrals of motion emerging from the second order of the
expansion are derived as: \be  {\mathbb I}_2 &=& Z_{11}^{(2)}-
\hat Z_{11}^{(2)} + {\mathrm f}_2^+ + {\mathrm f}_2^- = {4\beta^3
\over 3 m^2 }({\cal P}_2^{(b)} + i\sqrt 3 {\cal H}_2^{(b)}) , \non\\
\tilde {\mathbb I}_2 &=& Z_{22}^{(2)} - \hat Z_{22}^{(2)} +
{\mathrm h}_2^+ + {\mathrm h}_2^- = {4 \beta^3 \over 3 m^2 }({\cal
P}_2^{(b)} - i\sqrt 3 {\cal H}_2^{(b)})\ee where \be  {\cal
P}_2^{(b)} &=& {1 \over 2} \int_{-L}^0 dx \Big ({8\over \beta^3}(
\hat a \hat b \hat c- abc ) +{8\over \beta^3} (bc' +\hat b \hat
c') +{m^2 \over 2\beta^3}(\gamma_1(c-\hat c) +\gamma_2(a-\hat a)
+\gamma_3 (b-\hat b)) \Big )  \non\\
&+& {m^2\over 4 \beta^3} \Big ( \gamma_1(0) - \gamma_2(0) \Big ) +
{3m^2 \over 4 \beta^3}  e^{\beta \alpha_2 \cdot \Phi(0)}- {3m^2
\over 8 \beta^3}e^{-4 i \xi^+} e^{-2\beta \alpha_3 \cdot \Phi(0)}
\non\\ &+&{3m \over 2 \beta^3}e^{-2 i \xi^+}  e^{-\beta \alpha_3
\cdot \Phi(0)} \Big ( c(0) + \hat c(0) \Big  ) + {2 \over \beta^3}
\Big (\hat b'(0)-b'(0) \Big ) + {2\over \beta^3} \Big (b^2(0) +
{\hat b}^2(0) \Big )\non\\ &+& {m^2\over 4 \beta^3} \Big
(\gamma_2(-L) - \gamma_1(-L) \Big )+ {3m^2 \over 4 \beta^3} e^{\beta
\alpha_1 \cdot \Phi(-L)} -{3m^2 \over 8
\beta^3}e^{-4i\xi^-}  e^{-2 \beta \alpha_3 \cdot \Phi(-L)}\non\\
&+& {3m \over 2 \beta^3} e^{-2i \xi^-}  e^{-\beta \alpha_3 \cdot
\Phi(-L)} \Big (a(-L) + \hat a (-L) \Big )  +{2 \over \beta^3}
\Big ( b'(-L) - \hat b' (-L) \Big ) + {1\over \beta^3
}\Big (\hat b^2(-L) +b^2(-L)  \Big ) \non\\
\ee \be {\cal H}_{2}^{(b)} &=& {1 \over 2} \int_{-L}^L dx \Big
({8\over \beta^3}( abc +\hat a \hat b \hat c) -{8\over \beta^3}
(bc' -\hat b \hat c') -{m^2 \over 2\beta^3}(\gamma_1(c+\hat c)
+\gamma_2(a+\hat a) +\gamma_3 (b+\hat b)) \Big ) \non\\ &+& {3 m
\over 2\beta^3 } e^{-2 i \xi^+}  e^{-\beta \alpha_3 \cdot \Phi(0)}
\Big (\hat c(0) - c(0)\Big ) - {2 \over \beta^3} \Big (b'(0) +
\hat b'(0)\Big
) + {2\over \beta^3 }\Big ( b^2(0)  -\hat b^2(0) \Big ) \non\\
&+& {3 m \over 2\beta^3 } e^{-2i \xi^{-}}  e^{-\beta \alpha_3
\cdot \Phi(-L)}\Big ( \hat a(-L)- a(-L) \Big ) + {2 \over \beta^3}
\Big (b'(-L)+\hat b'(-L) \Big )+ {1\over \beta^3} \Big (
b^2(-L)-\hat b^2(-L) \Big ). \non\\ \ee Higher integrals of
motions are naturally obtained from the higher order expansion of
the open transfer matrix but we shall not further pursue this
point here. Notice that both ${\cal H}_k^{(b)}$ and ${\cal
P}_k^{(b)}$ are conserved quantities contrary to what happens in
the SNP case analyzed in the previous section, where only ${\cal
H}_k^{(b)}$ are conserved. This is another basic qualitative
difference between the two types of boundary conditions. Note that
from the deduced integrals of motion certain sets of equations of
motion are entailed. In particular the equations of motion arise
from the following equations: \be && {\partial \phi_i(x,t)\over
\partial t} = \Big \{{\cal H}_{1}^{(b)}(0,-L),\ \phi_i(x,t) \Big \}, ~~{\partial
 \pi_i(x,t)\over \partial t} = \Big \{{\cal H}_1^{(b)}(0,-L),\
\pi_i(x,t) \Big \}, \non\\ && -L \leq x \leq 0, ~~~~~i \in \{1,
\ldots, n \}. \label{eqmo} \ee A detailed discussion on the
associated equations of motion and the relevant boundary Lax pairs
systematically constructed along the lines described in
\cite{avandoikou} will be presented in a forthcoming publication.

As in the analysis of the preceding section for the classical
twisted Yangian (SNP) we may as well consider dynamical boundary
conditions in the SP case. Specifically, one can assume a generic
--dynamical-- representation of the underlying classical reflection
algebra (\ref{refc}) defined as \cite{sklyanin}: \be {\mathbb
K}(\lambda) = L(\lambda -\Theta)\ K(\lambda) \otimes {\mathbb I}\
L^{-1}(-\lambda -\Theta) \ee where $K$ is a $c$-number solution of
the classical reflection algebra \cite{dvg}, and $L$ is any solution
of (\ref{basic}). Again the extra boundary degrees of freedom are
incorporated in $L$. A more detailed analysis of such boundary
conditions in the ATFT frame will be presented elsewhere (see
similar analysis for the sine-Gordon and the vector NLS models in
\cite{bade} and \cite{nls} respectively).

\section{Discussion}

An exhaustive study of the integrable boundary conditions in
$A_n^{(1)}$ ATFT was presented by systematically deriving the
associated local integrals of motion. The key point in our
analysis is the extraction of the local integrals of motion
directly from the transfer matrix asymptotic expansion, and there
is no conjecture involved as far as their structure is concerned.
The systematic derivation of the boundary integrals of motion
starting from the underlying algebra gives rise to two distinct
types of boundary conditions associated to the reflection algebra
and $q$-twisted Yangian.

Noticeably the SP boundary conditions are absent in the analysis
presented in \cite{durham} mainly because of the a priori strong
constraints imposed upon the structure of the boundary conserved
local quantities. In \cite{durham} quantities of the type ${\cal
P}_k$ were a priori disregarded as non conserved --this is true
however only for the sine(sinh)-Gordon model ($A_1^{(1)}$)--
whereas as we see in the present investigation these objects play
a key role in distinguishing the two types of boundary conditions!
Although sine-Gordon is the prototype model of the class under
consideration an `imitation' of its boundary behavior by the
higher members of the hierarchy could be quite misleading. This is
primarily due to the fact that the sine-Gordon is a self-conjugate
model --soliton and anti-soliton are equivalent entities-- and as
such it has a very peculiar boundary behavior that cannot be
naively generalized to higher $A_n^{(1)}$ ATFT.

One of the basic differences between the two types of boundary
conditions is that in the SP case the number of integrals of
motion is `double' compared to the SNP ones. This phenomenon not
only indicates a qualitatively different behavior of the model as
far as the boundaries are concerned, but also leads to a
modification of the bulk behavior altogether (see also e.g.
\cite{nls}). The `duplication' of the local integrals of motion in
the SP case seems to persist to higher orders --we checked
explicitly up to third order. More precisely, let ${\cal Q}_k,\
{\cal Q}_{-k}$ be the local integrals of motion of the $A_n^{(1)}$
ATFT on the full line, then the boundary conserved quantities for
each type of boundary conditions are provided by: \be && {\cal
Q}^{(b)}_k  = {\cal Q}_k + {\cal Q}_{-k} + {\cal B}_k
~~~~~\mbox{for SNP} \non\\
&& {\cal Q}_{k}^{\pm (b)}  = {\cal Q}_k \pm {\cal Q}_{-k} + {\cal
B}^{\pm}_k ~~~~~\mbox{for SP} \ee ${\cal B}_k,\ {\cal B}^{\pm}_k$
are the relevant boundary terms. In the SNP case only the integrals
of motion where the bulk part is provided by the sum of ${\cal
Q}_k,\ {\cal Q}_{-k}$ survive, while in SNP both sums and
differences provide local conserved quantities, i.e. each one of
${\cal Q}_{\pm k}$ (with appropriate boundary terms) is a conserved
quantity. Moreover in the SNP case no free parameters appear in the
integrals of motion due to fact that the corresponding $c$-number
$K$-matrices contain no free parameters. However in the SP case, as
anticipated, the relevant integrals of motion depend on the
parameters $\xi^{\pm},\ l^{\pm}$. It is worth stressing that in the
context of integrable spin chains the parameters $\xi^{\pm},\
l^{\pm}$ explicitly appear in the corresponding Hamiltonian as well
as in the associated symmetry of the model. More precisely, it was
shown in \cite{done} that the rational open spin chain with diagonal
boundary conditions associated to integers $l^{\pm} = l$ is $gl_{l}
\otimes gl_{n+1-l}$ invariant and ${\cal U}_q(gl_l)\otimes {\cal
U}_q(gl_{n+1-l})$ invariant in the trigonometric case, relevant to
the ATFT theories. Recall that the ${\cal U}_q(gl_{n+1})$ spin chain
maybe thought of as an integrable lattice version of the $A_n^{(1)}$
ATFT in the same logic that the critical XXZ spin chain may be seen
as the lattice version of the sine-Gordon model.

There exist various studies concerning the underlying symmetry
algebras when non-trivial integrable boundary conditions are
present. Specifically, the symmetry algebra in the context of ATFT
with SNP boundary conditions --being a twisted algebra-- was
investigated in \cite{dema}, while extensive studies on the
underlying algebras in integrable spin chains with both types of
boundary conditions are presented in \cite{doikoun, doikouy}. An
analysis in the spirit of \cite{dema, bela} would provide the
non-local integrals of motion forming the exact symmetry algebra
in the SP case, however this will be presented in a separate
publication (see a relevant analysis in the quantum case in
\cite{doikoun}).

Another intriguing point associated to the `folding' of integrals
of motion is the possible folding of the classical counterparts of
Bethe ansatz equations in the SNP case emerging from the solution
of the spectral problem \cite{ft, babevi}. Although folding of
Bethe anastz equations has been reported so far only in isotropic
examples we conjecture that it should also occur in models
associated to trigonometric $R$-matrices. In general the structure
of Bethe ansatz is immediately linked to the underlying algebra,
therefore a folding of the associated algebra --and the
corresponding Dynkin diagrams --would be reflected to the
structure of the Bethe equations. Extensive studies on the folding
of the Bethe equations and the relevant Dynkin diagrams are
presented in \cite{ann1, ann2, doikousnp}.

In a more physical frame this would be translated to a folding of
the associated exact boundary $S$-matrices. Notwithstanding
boundary $S$-matrices were extracted in \cite{done} in the SP
case, the derivation of boundary $S$-matrices in the SNP case is
still an open question to date in the general case (see e.g.
\cite{gand}). Having said this the derivation of the Bethe ansatz
equations for trigonometric spin chains with SNP boundary
conditions, and the associated boundary $S$-matrices will provide
significant information at both physical and algebraic level.

The next natural step would be to identify the relevant boundary
Lax pairs for both types of boundary conditions along the lines
described in \cite{avandoikou}. A comparison with the Lax pair
constructed based on a set of postulates in \cite{durham} will be
especially illuminating. In the SNP case the entailed Lax pair
should coincide with that found in \cite{durham}, whereas the Lax
pair in the SP case will be of a novel from. Generalization of our
results for any $A_n^{(1)}$ ($n >
 1$) ATFT will be also presented
in a separate publication. Finally, a similar exhaustive analysis
regarding principal chiral models (partial results maybe found in
\cite{dema2}) will be particularly relevant especially bearing in
mind the physical significance of a specific super-symmetric
principal chiral model within the AdS/CFT correspondence
\cite{polch, kaza}.

\noindent{\bf Acknowledgments:} I am indebted to J. Avan for
useful comments. I wish to thank INFN, Bologna Section, and
University of Bologna for kind hospitality. This work was partly
supported by INFN, Bologna section, through grant TO12.

\appendix

\section{Appendix}

In this appendix we provide explicit expressions of the simple
roots and the Cartan generators for $A_{n}^{(1)}$ \cite{georgi}.
The vectors $\alpha_{i} = (\alpha_{i}^{1} \,, \ldots \,,
\alpha_{i}^{n})$ are the simple roots of the Lie algebra of rank
$n$ normalized to unity $\alpha_{i} \cdot \alpha_{i} = 1$, i.e.
\be \alpha_{i} = \Bigl(0 \,,  \ldots \,, 0 \,, -\sqrt{i-1\over 2i}
\,, \stackrel{\stackrel{i^{th}}{\downarrow}} {\sqrt{i+1\over 2i}}
\,, 0 \,,  \ldots \,, 0 \Bigr), ~~~~i \in \{1, \ldots n \} \ee
Also define the fundamental weights $\mu_{k} = (\mu_{k}^{1} \,,
\ldots \,, \mu_{k}^{n}) \,, \quad  k = 1 \,, \ldots \,, n $ as
(see, e.g., \cite{georgi}). \be \alpha_{j} \cdot \mu_{k} = {1\over
2} \delta_{j,k} \,. \label{important} \ee The extended (affine)
root $a_{n+1}$  is provided by the relation \be \sum_{i=1}^{n+1}
a_i =0. \ee We give below the Cartan-Weyl generators in the
defining representation: \be E_{\alpha_{i}} &=& e_{i\ i+1} \,,
\qquad E_{-\alpha_{i}} = e_{i+1\ i} \,, \qquad E_{\alpha_n} = -
e_{n+1\ 1} \,,
\qquad E_{-\alpha_n} = - e_{1\ n+1}\non \\
H_{i} &=& \sum_{j=1}^{n} \mu_{j}^{i} (e_{j j} -e_{j+1\ j+1}) \,,
\qquad i = 1 \,, \ldots \,, n \label{cartan/weyl/basis} \ee For
$A_{2}^{(1)}$ in particular we have: \be \alpha_1 = (1,\ 0),
~~~\alpha_2 = (-{1\over 2},\ {\sqrt 3 \over 2}),~~~~\alpha_3 =
(-{1\over 2},\ -{\sqrt 3 \over 2}) \ee define also the following
$3\times 3$ generators \be E_{1} = E_{-1}^t = e_{12}, ~~~~E_2 =
E^t_{-2} = e_{23}, ~~~~E_3= E^t_{-3} = -e_{31} \ee where we define
the matrices $e_{ij}$ as $(e_{ij})_{kl} = \delta_{ik}\
\delta_{jl}$. The diagonal Cartan generators $H_{1,2}$ are \be H_1
={1\over 2} (e_{11} -e_{22}), ~~~~H_2={1\over 2 \sqrt 3}
(e_{11}+e_{22} -2 e_{33}) \ee

\section{Appendix}

From the formulas (\ref{form}), (\ref{zz0}) the matrices
$W^{(k)},\ \hat W^{(k)},\ Z^{(k)},\ \hat Z^{(k)}$ may be
determined. In particular, we write below explicit expressions of
these matrices for the first orders. \be && W^{(0)} = \hat
W^{(0)}= \left(
\begin{array}{ccc}
0                    & e^{{i \pi \over 3}}  & 1 \\
e^{{i \pi \over 3}}  & 0                    & -1 \\
e^{{2i \pi \over 3}} & e^{-{i \pi \over 3}} & 0\\ \end{array} \right), \non\\
&& {m\over 4} W^{(1)}= \left( \begin{array}{ccc}
0       & e^{{2i \pi \over 3}} a  & c \\
-a      & 0                       & b \\
e^{{i \pi \over 3}}c  & -b        & 0\\ \end{array} \right), ~~~~{m\over 4}
\hat W^{(1)} = \left( \begin{array}{ccc}
0       & -\hat b  & -\hat a \\
-e^{-{i \pi \over 3}}\hat b      & 0                       & -\hat c \\
\hat a  & -e^{{i \pi \over 3}}\hat c        & 0\\ \end{array}
\right). \ee The higher order quantities are more complicated and
we give the matrix entries below for $W^{(2)},\ \hat W^{(2)}$
(define also, $~\zeta = {4 \over m}$): \be && W_{12}^{(2)} =
{1\over 3} (-2 \gamma_3 + \gamma_1 +\gamma_2) + {\zeta^2 \over 3}
(2a' + b') + {\zeta^2 \over 3}
(-2a^2 -bc), \non\\
&& W^{(2)}_{21}= {e^{-{i\pi \over 3}} \over 3} (-2\gamma_3 + \gamma_1 +\gamma_2)
 + {\zeta^2 e^{-{i\pi \over 3}} \over 3} (a' - c') + {\zeta^2 e^{-{i\pi \over 3}}
\over 3}(c^2 -ab) \non\\
&& W_{13}^{(2)} =  {1\over 3}
(-2 \gamma_2+ \gamma_1 +\gamma_3) + {\zeta^2 \over 3} (-b' + c') +
{\zeta^2 \over 3}(b^2 -ac), \non\\
&& W_{31}^{(2)} =  {1\over 3}  (2 \gamma_2 - \gamma_1 -\gamma_3) +
{\zeta^2 \over 3} (-a' -2c') + {\zeta^2 \over 3}(2c^2 + a b), \non\\
&& W_{23}^{(2)} =  -{1\over 3}
(2 \gamma_1 -\gamma_2 - \gamma_3 ) + {\zeta^2 \over 3} (2b' + c') +
{\zeta^2 \over 3}(-2b^2 -ac) \non\\
&& W^{(2)}_{32}= -{e^{{i\pi \over 3}} \over 3}
(2 \gamma_1 - \gamma_2 - \gamma_3)
 + {\zeta^2 e^{{i\pi \over 3}} \over 3} (-a' +b') +
{\zeta^2 e^{{i\pi \over 3}} \over 3}(a^2 -b c) \ee and \be && \hat
W_{12}^{(2)} =  {e^{-{i\pi \over 3}} \over 3} (-2 \gamma_2 +\gamma_1
+ \gamma_3) + {\zeta^2e^{-{i\pi \over 3}} \over 3} (\hat  b' - \hat
c') + {\zeta^2 e^{-{i\pi \over 3}}
\over 3}(\hat c^2 - \hat a \hat b), \non\\
&& \hat W^{(2)}_{21}= {1 \over 3} (-2 \gamma_2 + \gamma_1 +\gamma_3)
 + {\zeta^2  \over 3} (2\hat b' +\hat a')+ {\zeta^2 \over 3}
(-2 \hat b^2 -\hat a \hat c) \non\\
&& \hat W_{13}^{(2)} =  -{1\over 3} (-2 \gamma_1  +\gamma_3 +\gamma_2) - {\zeta^2 \over 3}
(2\hat a' + \hat c') +
{\zeta^2 \over 3} (2\hat a^2 +\hat b \hat c), \non\\
&& W_{31}^{(2)} =  {e^{{i\pi \over 3}}\over 3}  (2 \gamma_1 -\gamma_2-\gamma_3) +
{\zeta^2 e^{{i\pi \over 3}} \over 3} (\hat b' -\hat a') + {\zeta^2 e^{{i\pi \over 3}}
\over 3}(-\hat b^2 +\hat a \hat c), \non\\
&& \hat W_{23}^{(2)} =  -{1\over 3}
(-2 \gamma_3  +\gamma_2 + \gamma_1 ) + {\zeta^2 \over 3} (\hat a' -\hat c') + {\zeta^2 \over 3}
(-\hat a^2 + \hat b \hat c) \non\\
&& \hat W^{(2)}_{32}= {1 \over 3} (-2 \gamma_3 +\gamma_1 + \gamma_2 )
 + {\zeta^2  \over 3} (\hat b' + 2 \hat c') + {\zeta^2  \over 3}(-2 \hat c^2 -
\hat a \hat b) \ee where the prime denotes derivative with respect
to $x$, also $a,\ b,\ c,$ and $\gamma_i$ are defined in
(\ref{def}) and have the following explicit forms: \be && a =
{\beta \over 2} ({\theta_1 \over 2} + {\theta_2 \over 2 \sqrt 3}
), ~~~~b = {\beta \over 2} (-{\theta _1 \over 2}+ {\theta_2 \over
2 \sqrt 3}),
~~~~c = -{\beta \over 2}  {\theta_2 \over  \sqrt 3}, \non\\
&& \gamma_1 = e^{\beta \phi_1}, ~~~~\gamma_2 = e^{\beta(-{1 \over 2}\phi_1 + {\sqrt 3 \over 2} \phi_2)},
 ~~~~~\gamma_3=e^{\beta(-{1 \over 2}\phi_1 - {\sqrt 3 \over 2} \phi_2)}. \label{def2} \ee
Moreover using the expressions above and (\ref{zz0}) we have: \be
&& {d Z_{11}^{(1)}\over dx} = {e^{-{i\pi \over 3}}\over 3} {m
\over 4}(\gamma_1 +\gamma_2 + \gamma_3)  +{\zeta e^{-{i\pi \over
3}}\over 3} (a' -c') +{\zeta e^{-{i\pi \over 3}} \over 6} (a^2
+b^2 +c^2) \non\\ && {d Z_{22}^{(1)} \over dx} = {e^{{i\pi \over
3}} \over 3} {m \over 4}(\gamma_1 +\gamma_2 + \gamma_3)  +{\zeta
e^{{i\pi \over 3}} \over 3} (b' -a') +{\zeta e^{{i\pi \over
3}}\over 6} (a^2 +b^2 +c^2) \non\\ && { d Z_{33}^{(1)} \over dx} =
-{1\over 3} {m \over 4}(\gamma_1 +\gamma_2 + \gamma_3)  -{\zeta
\over 3} (c' -b') -{\zeta \over 6}
(a^2 +b^2 +c^2) \non\\
&& {d \hat Z_{11}^{(1)} \over dx} = {e^{{i\pi \over 3}} \over 3}
{m \over 4}(\gamma_1 +\gamma_2 + \gamma_3) -{\zeta e^{{i\pi \over
3}} \over 3} (\hat b' -\hat a') +{\zeta e^{{i\pi \over 3}}\over 6}
(\hat a^2 +\hat b^2 +\hat c^2)
\non\\
&& {d \hat Z_{22}^{(1)} \over dx} = {e^{-{i\pi \over 3}}\over 3}
{m \over 4}(\gamma_1 +\gamma_2 + \gamma_3)  +{\zeta e^{-{i\pi
\over 3}}\over 3} (\hat b' - \hat c') +{\zeta e^{-{i\pi \over 3}}
\over 6} (\hat a^2 +\hat b^2 +\hat c^2) \non\\ && {d \hat
Z_{33}^{(1)}\over dx} = -{1\over 3} {m \over 4}(\gamma_1 +\gamma_2
+ \gamma_3) +{\zeta \over 3} (\hat a' -\hat c') -{\zeta \over 6}
(\hat a^2 +\hat b^2 + \hat c^2) \label{z1} \ee Finally we report
$Z_{ii}^{(2)},\ \hat Z_{ii}^{(2)}$: \be && {d Z_{11}^{(2)} \over
dx} = {e^{{i\pi \over 3}}\over 3} \Big (\gamma_2' -\gamma_3' -
\zeta^2 (c''- c^{2'}) + \zeta^2 ca' +(\gamma_1 c + \gamma_2 a +
\gamma_3 b) - \zeta^2 abc \Big ) \non\\
&& {d Z_{22}^{(2)} \over dx} = {e^{-{i\pi \over 3}}\over 3} \Big
(-\gamma_1' + \gamma_3' - \zeta^2 (a''- a^{2'}) + \zeta^2 a b'
+(\gamma_1 c + \gamma_2 a +
\gamma_3 b) - \zeta^2 abc \Big ) \non\\
&& {d Z_{33}^{(2)} \over dx} = {1\over 3} \Big (-\gamma_1' +
\gamma_2' + \zeta^2 (b''- b^{2'}) - \zeta^2 b c' -(\gamma_1 c +
\gamma_2 a + \gamma_3 b) +
\zeta^2 abc \Big ) \non\\
&& {d \hat Z_{11}^{(2)} \over dx} = {e^{-{i\pi \over 3}}\over 3}
\Big (-\gamma_1' +\gamma_2' - \zeta^2 (\hat b''- \hat b^{2'}) +
\zeta^2  \hat b \hat a' +(\gamma_1 \hat c + \gamma_2 \hat a +
\gamma_3 \hat b) - \zeta^2
\hat a \hat b \hat c \Big ) \non\\
&& {d \hat Z_{22}^{(2)} \over dx} = {e^{{i\pi \over 3}}\over 3}
\Big (-\gamma_2' +\gamma_3' - \zeta^2 (\hat c''- \hat c^{2'}) +
\zeta^2  \hat c \hat b' +(\gamma_1 \hat c + \gamma_2 \hat a +
\gamma_3 \hat b) - \zeta^2 \hat a \hat b \hat c \Big ) \non\\
&& {d \hat Z_{33}^{(2)} \over dx} ={1 \over 3} \Big (-\gamma_1'
+\gamma_3' + \zeta^2 (\hat a''- \hat a^{2'}) - \zeta^2 \hat a \hat
c' -(\gamma_1 \hat c + \gamma_2 \hat a + \gamma_3 \hat b) + \zeta^2
\hat a \hat b \hat c \Big ). \label{z2} \ee

\section{Appendix}

We present here the boundary contributions in the expansion of the
classical open transfer matrix for both types of boundary
conditions: \\
\\
{\bf SNP boundary conditions}: Recall that in this case the
expansion of the generating function of the local integrals of
motion is given in (\ref{expan2}). After some tedious algebra we
obtain for the boundary terms: \be {\mathrm f}_0 &=& \ln (3\bar g),
~~~~~~{\mathrm f}_1 ={1\over 3\bar g} \Big (e^{{\beta \over 2}
\alpha_3 \cdot \Phi(0)}f_{31} - e^{{\beta \over 2} \alpha_2 \cdot
\Phi(0)}f_{23} -   e^{{\beta \over 2} \alpha_1 \cdot
\Phi(0)}f_{12} \Big ) + {\zeta \over 3} \Big (c(0)-b(0) +\hat c(0) - \hat a(0) \Big ) \non\\
{\mathrm f}_2 &=& -{1\over 3 \bar g} \Big (f_{21} e^{-{\beta \over
2} \alpha_1 \cdot \Phi(0)} + f_{32} e^{-{\beta \over 2}\alpha_2
\cdot
\Phi(0)} -f_{13} e^{-{\beta \over 2} \alpha_3 \cdot \Phi(0)}\Big ) \non\\
&-& {\zeta \over 3 \bar g} \Big (f_{12}e^{{\beta \over 2} \alpha_1
\cdot \Phi(0)} (c(0) +\hat c(0) ) -f_{23} e^{{\beta \over 2}
\alpha_2 \cdot \Phi(0)} b(0) + f_{31} e^{{\beta \over 2} \alpha_3
\cdot
\Phi(0)} \hat a(0) \Big ) - {\zeta^2 \over 3} \Big (\hat a(0) c(0) + b(0) \hat c(0)\Big ) \non\\
&+& {1\over 3} \Big ( (2 \gamma_1(0) -\gamma_2(0) -\gamma_3(0)) -
\zeta^2 (\hat a'(0) + b'(0)) + \zeta^2 (\hat a^2(0) + b^2(0)) \Big )\non\\
&-& {1\over 2} \Big ( {1 \over 3 \bar g} (e^{{\beta \over 2}
\alpha_3 \cdot \Phi(0)}f_{31} - e^{{\beta \over 2} \alpha_2 \cdot
\Phi(0)}f_{23} - e^{{\beta \over 2} \alpha_1 \cdot \Phi(0)}f_{12} )
+ {\zeta \over 3}(c(0)-b(0) +\hat c(0) - \hat a(0)) \Big )^2.
\label{boundary} \ee
\\
{\bf SP boundary conditions}: We shall need for our purposes here
the asymptotics of $K^{\pm}$ as $|u| \to \infty$: \be K^+(|u| \to
\infty,\ \xi^+) & \sim & e_{33} -{ e^{-2 i \xi^+}
\over u} e_{11} + {1 \over u^2} e_{22} + {\cal O}(u^{-3}) \non\\
K^{-}(|u| \to \infty,\ \xi^-) & \sim & e_{11} - {e^{-2i \xi^-}
\over u} e_{33} + {1\over u^2} e_{22} + {\cal O}(u^{-3}). \ee Then
from the expansion of the boundary terms in (\ref{bound2}),
(\ref{expp2}) we obtain the following explicit quantities: \be
{\mathrm f}^{+}_0 &=& {\mathrm h}_0^{+} =\ln [{\Omega_{33}^2(0)
\over 3}] , ~~~~{\mathrm f}^+_1 = -\zeta e^{i\pi \over 3} \hat
b(0) + \zeta e^{-{i\pi \over 3}} c(0) - e^{-2 i \xi^+} e^{-\beta
\alpha_3 \cdot \Phi(0) }, \non\\
{\mathrm f}^+_2 &=& \Big \{ \Omega_{22}^2(0) \Omega_{33}^{-2}(0) -
{e^{-4 i\xi^+} \over 2}\Omega_{11}^4(0) \Omega_{33}^{-4}(0)  +
{\zeta e^{-2 i\xi^+} \over 2} \Omega_{11}^2(0) \Omega_{33}^{-2}(0) \Big (c(0) +\hat c (0)\Big ) \non\\
& -&{1 \over 6} \Big ( 2\gamma_2(0) -\gamma_1(0) -\gamma_3(0)\Big )
-{\zeta^2 \over 6} \Big (b'(0) -c'(0) \Big ) + \zeta^2 \Big
(-{c^2(0) \over 6 } + {a^2(0)
\over 12}+{b^2(0) \over 12} + {\hat b^2(0) \over 4 } \Big ) \Big \}\non\\
& +& i \sqrt 3 \Big \{ {\zeta e^{-2i\xi^+} \over 2}
\Omega_{11}^2(0) \Omega_{33}^{-2}(0) \Big (\hat c(0) - c(0)\Big )
-{1 \over 6} \Big( 2\gamma_2(0) - \gamma_1(0)- \gamma_3(0) \Big )
-{\zeta^2 \over 6}\Big (b'(0) -c'(0)\Big ) \non\\ &+& \zeta^2 \Big
(-{c^2(0) \over 6} + { a ^2(0) \over 12} + {b^2(0) \over 12}
-{\hat b^2(0) \over 4}  \Big ) \Big \} \non\\ {\mathrm h}_1^+ &=&
- \zeta e^{i\pi \over 3} b(0) - e^{-2i \xi^+}
e^{-\beta \alpha_3 \cdot \Phi(0) } \non\\
 {\mathrm h}^+_2 &=& \Big\{ \Omega_{22}^2(0) \Omega_{33}^{-2}(0) - {e^{-4
i \xi^+} \over 2 } \Omega_{11}^4(0) \Omega_{33}^{-4}(0) + {\zeta
e^{-2 i \xi^+} \over 2} \Omega_{11}^2(0) \Omega_{33}^{-2}(0) \Big (
c(0) + \hat c(0)\Big  ) + {1 \over 3} \Big (\gamma_1(0) -
\gamma_2(0)\Big) \non\\ & + &{\zeta^2 \over 6} \Big (\hat b'(0)
-\hat c'(0) -b'(0) +a'(0) \Big ) + \zeta^2 \Big ({\hat a^2(0) \over
12 }+ {\hat b^2(0) \over 12} + {\hat c^2(0) \over 12} + {b^2(0)
\over 6} - {a^2(0) \over 12} -{c^2(0) \over 12} \Big )\Big \} \non\\
& +&i\sqrt 3 \Big \{ {\zeta e^{-2i \xi^+}\over 2} \Omega_{11}^2(0)
\Omega_{33}^{-2}(0) \Big (c(0) -\hat c(0)\Big ) + {1\over 6}\Big (2
\gamma_3(0) -\gamma_1(0) -\gamma_2(0)\Big ) \non\\ & +& {\zeta^2
\over 6}\Big (\hat b'(0) -\hat c'(0) +b'(0) - a'(0) \Big ) + \zeta^2
\Big ( {\hat a^2(0) \over 12 }+ {\hat b^2(0) \over 12} + {\hat
c^2(0) \over 12} - {b^2(0) \over 6} + {a^2(0) \over 12} +{c^2(0)
\over 12} \Big ) \Big \} \ldots \label{bout} \ee Similar expressions
are obtained for ${\mathrm f}_n^-,\ {\mathrm h}_n^-$: \be {\mathrm
f}_0^- &=& {\mathrm h}_0^- = \ln [{\Omega^{-1}_{11}(-L) \over 3 }],
~~~~~~{\mathrm f}_1^- = \zeta e^{-{i\pi \over 3}} a
(-L) - e^{-2 i \xi^-} e^{-\beta \alpha_3 \cdot \Phi(-L) }  \non\\
 {\mathrm f}_2^- &=& \Big \{ \Omega_{11}^2(-L) \Omega_{22}^{-2}(-L)
- {e^{-4i\xi^-} \over 2} \Omega_{11}^4(-L) \Omega_{33}^{-4}(-L)
+{\zeta e^{-2i \xi^-} \over 2}
\Omega_{11}^2(-L) \Omega_{33}^{-1}(-L) \Big (a(-L) +\hat a (-L) \Big ) \non\\
& + &{1 \over 6} \Big (2\gamma_2(-L) - \gamma_1(-L) - \gamma_3(-L)
\Big ) + \zeta^2  \Big ( -{a'(-L)\over 18} - {4c'(-L)\over 18} + {2b'(-L)\over 18}\Big ) \non\\
& + &\zeta^2 \Big ({2 c^2(-L) \over 12} -{a^2(-L) \over 12
}+{2b^2(-L) \over 12} \Big )\Big \} \non\\
& + &i \sqrt{3} \Big \{{\zeta e ^{-2i\xi^-} \over
2}\Omega_{11}^2(-L)\Omega_{33}^{-2}(-L) \Big (\hat a(-L) - a(-L)
\Big ) + {1\over 6} \Big ( 2\gamma_2(-L)
-\gamma_1(-L)-\gamma_3(-L)\Big ) \non\\ & +&\zeta^2 \Big (-{a'(-L)
\over 18} -{4 c'(-L) \over 18} +{2b'(-L) \over 18}\Big )+ \zeta^2
\Big ( {2c^2(-L) \over 12 } -{a^2(-L) \over 12} + {2b^2(-L) \over 12} \Big ) \Big \} \non\\
 {\mathrm h}^-_1 &=& -\zeta e ^{- {i \pi \over 3}} \hat b(-L) -
e^{-2 i\xi^-} e^{-\beta \alpha_3 \cdot \Phi(-L) } \non\\ {\mathrm
h}_2^- &=& \Big \{ \Omega_{11}^2(-L) \Omega_{22}^{-2}(-L) -
{e^{-4i\xi^-} \over 2} \Omega_{11}^4(-L) \Omega_{33}^{-4}(-L)
+{\zeta e^{-2i \xi^-} \over 2}
\Omega_{11}^2(-L) \Omega_{33}^{-1}(-L) \Big (a(-L) +\hat a (-L) \Big ) \non\\
& -& {1 \over 2} \Big (\gamma_1(-L) - \gamma_2(-L)
\Big ) + \zeta^2  \Big ( -{\hat b'(-L)\over 6} + {\hat c'(-L)\over 6}-
{a'(-L)\over 6}  + {b'(-L)\over 6}\Big ) \non\\
&+ &\zeta^2 \Big ( - {\hat c^2(-L) \over 12}- {\hat a^2(-L) \over
12 } +{\hat b^2(-L) \over 6}+ {c^2(-L) \over 12} +{a^2(-L) \over
12} + {b^2(-L) \over 12 }\Big )\Big \} \non\\
& +& i \sqrt{3} \Big \{{\zeta e ^{-2i\xi^-} \over
2}\Omega_{11}^2(-L)\Omega_{33}^{-2}(-L) \Big (-\hat a(-L) + a(-L)
\Big ) + {1\over 6} \Big ( -2\gamma_3(-L)
+\gamma_1(-L)+\gamma_2(-L)\Big ) \non\\ & +& \zeta^2 \Big (-{\hat
b'(-L) \over 6} + {\hat c'(-L) \over 6} + {a'(-L) \over 6} -
{b'(-L) \over 6}\Big )\non\\ & +& \zeta^2 \Big (-{\hat c^2(-L)
\over 12 }-{\hat a^2(-L) \over 12 } + {\hat b^2(-L) \over 6} -
{a^2(-L) \over 12} - {b^2(-L) \over 12 } - {c^2(-L) \over 12} \Big
) \Big \}. \ee

\end{document}